\documentclass[fleqn]{article}
\usepackage{amsmath}
\usepackage{amssymb}
\usepackage{emlines2,bezier,amscd,amsmath}
\usepackage{amssymb}

\title{Structural instability of Friedmann-Robertson-Walker cosmological models}
\author{Sergey S. Kokarev}
\date{RSEC "Logos"\,, Yaroslavl, Respublikanskaya 80, r.22, e-mail: logos-center@mail.ru}

\begin{document}
\maketitle

\begin{abstract}
Cosmological singularity and asymptotic behaviour of scale factor
of generalized cosmological models are analyzed in respect of their structural stability.
It is shown, that cosmological singularity is structurally unstable
for the majority of models with barotropic perfect fluid with strong energy condition.
Inclusion of $\Lambda$-term extends the set of structurally stable
cosmological models.
\end{abstract}

\section{Introduction}

Relativistic Friedmann-Robertson-Walker cosmological models
(FRW-models) perform generally accepted current theoretical basis for
description of global structure and evolution of the Universe.
Remarkable property of these models is their simplicity: in
most situations we deal with one scale factor and equation of  state,
which has to be specified by some additional physical considerations.
Observational data allow in principle to correct some basic parameters of the models,
as well as
to specify properties of matter at different stages of evolution of the Universe and sometimes
they imply reconsideration of Einstein gravity in the context of some generalized
theories \cite{peldan,quint,odin1}. History of cosmology shows (and methodological principle
of eligibility prescribes) that correction of standard cosmological models
is realized mainly by sequence of their, in some sense, "small"\, modifications.
In this way we are  necessarily faced  with  situation which is well known
in qualitative differential equations theory
\cite{arnold}: some properties of original (non-modified) model
may "survive"\, after  "small"\, modification of the model, while
others may disappear.
In the former case the property is referred to as "rough"\, or structurally stable,
in the latter case --- as "thin"\, or structurally unstable.
As an example of such thin property of FRW-models we may consider cosmological
singularity.
By Hawking-Penrose theorem \cite{hawk} which has been proved within Einstein GR,
collapse is inevitably reached at some finite moment of time.
However, there are many exact cosmological solutions, obtained in frame of
generalized theories of gravitation (for example, involving scalar fields or non-riemannian objects
\cite{krech,fabris}), which describe evolution of the Universe without
singularity or with singularities of qualitatively  different kinds.
By general considerations consecutive cosmological model
should have structurally stable (with respect to possible modifications) basic properties.
In the opposite case any "small"\, work modification of the theory
will almost always lead to a change of cosmological paradigm.

In present paper we perform analysis of structural stability of
FRW-models.
We restrict ourselves by  investigations
of structural perturbations induced by   generalized
$f(R)$-theories of gravity and nonlinear equations of state.
Both of these generalizations of standard cosmology  have been actively investigated last decade in the context
of new cosmological observational data \cite{kremer,odin,caroll,song,mota,amen,odin2}.
We investigate structural stability of the two important properties of FRW-models:
cosmological singularity ($S$-stability) and asymptotic behaviour
of scale factor at $t\to\infty$ ($A$-stability).  Latter property
makes sense only for open and flat FRW-models.
These restrictions allow to reduce the problem to investigation of asymptotic solutions
to second order differential equation with variable coefficients.
In spite of quite different motivation of this work,
our analysis in some aspects reproduces and supplements results obtained in
\cite{song,bean,hu,sh1,sh2}.
In some important points we go to the conclusions made earlier in
\cite{sh1,sh2}.

In section \ref{stand} we outline class of standard models. In section
\ref{non} we describe those generalized models
which will be used for investigation of structural stability
of standard FRW-models.
Basic equation on structural perturbations and its analysis
are performed in section \ref{ge}. Section \ref{lambda} is devoted
to investigation of structural stability of standard FRW-models with
$\Lambda$-term. Here we apparently illustrate the idea of "structural security"\, of cosmological model.
In Conclusion we summarize and discuss all obtained results.

\section{Standard FRW-models} \label{stand}

Standard homogeneous isotropic cosmological models
can be described by metric of the following kind:
\begin{equation}\label{frid}
g=N^2(t)dt\otimes dt-a^2(t)(dr\otimes dr+\frac{\sin^2(kr)}{k^2}(d\theta\otimes d\theta+\sin^2\theta\,d\varphi\otimes
d\varphi)),
\end{equation}
where $N(t)$ is factor, defining gauge of cosmological time,
$a(t)$ is scale factor, describing cosmological evolution of space lengths,
$k$ is curvature constant parameter, $r$ is radial coordinate,  $\theta$ and
$\varphi$ are angle coordinates. The parameter $k$ has dimension of inverse length and
can be real, imaginary or null. In the first case we deal with
closed cosmological models, in the second case --- with open cosmological models,
in the third case --- with flat ones.
In what follows all formulas will have universal sense for all
possible values of parameter $k.$ The case $k=0$ sometimes will require
rather simple limit procedure: $k\to0.$

Standard calculations give the following nonzero components of
mixed Ricci tensor:
\begin{equation}\label{riccg}
R^0_0=\frac{3}{N^2}\left(\frac{\dot N\dot a}{Na}-\frac{\ddot
a}{a}\right);\quad
R^1_1=R^2_2=R^3_3=\frac{1}{N^2}\left(\frac{\dot N\dot a}{Na}-\frac{\ddot
a}{a}-2\frac{\dot a^2}{a^2}\right)-2\frac{k^2}{a^2}.
\end{equation}
Scalar curvature is expressed by the formula\footnote{Hereafter Greek indexes $\alpha,\beta,\gamma$
denote space-time components of 4-tensors,
Latin indexes $i,j,k$ --- their space components.}:
\begin{equation}\label{scalg}
R\equiv R^\alpha_\alpha=\frac{6}{N^2}\left(\frac{\dot N\dot a}{Na}-\frac{\ddot
a}{a}-\frac{\dot a^2}{a^2}\right)-6\frac{k^2}{a^2}.
\end{equation}

It is commonly accepted, that source of gravity in cosmology is
barotropic perfect fluid with energy-momentum tensor of the kind:
\begin{equation}\label{energ}
T=(\varepsilon+p)u\otimes u-p g,
\end{equation}
where $u=N dt$ --- 1-form of 4-velocity of matter, defining
comoving reference frame,
$\varepsilon$ and $p$ ---
scalars of energy density and pressure of the matter, connected with each other
by barotropic equation of state:
\begin{equation}\label{eqstat}
p=\alpha\varepsilon,\quad\text{where}\quad\alpha=\text{const}.
\end{equation}
Apart from perfect fluid for the more generality of our formulas
we include into the standard models $\Lambda-$term (which,
excepting section \ref{lambda}, we shall assume zero).
Einstein equations:
\begin{equation}\label{eing}
G^\alpha_\beta\equiv R^\alpha_\beta-\frac{1}{2}\delta^\alpha_\beta
R=\kappa T^\alpha_\beta+\Lambda\delta^\alpha_\beta,
\end{equation}
derived from Einstein-Hilbert  variational principle for the action
($S_{\text{m}}$ --- action for matter,
$\kappa=8\pi G/c^4$ --- Einstein gravitational constant, $\ast$ ---
standard dual conjugation):
\begin{equation}\label{act0}
S[g]=-\frac{1}{2\kappa}\int \ast (R+2\Lambda)+S_{\text{m}},
\end{equation}
take the form of basic dynamical equations:
\begin{equation}\label{eq10}
\frac{3\dot a^2}{N^2 a^2}+\frac{3k^2}{a^2}=\kappa \epsilon+\Lambda;\quad
\frac{1}{N^2}\left(-\frac{2\dot N\dot a}{Na}+2\frac{\ddot
a}{a}+\frac{\dot a^2}{a^2}\right)+\frac{k^2}{a^2}=-\kappa p+\Lambda.
\end{equation}

For the simplicity sake we accept conformal gauge of cosmological
metric (\ref{frid}), which is defined by condition: $N=a.$
Also we go to units, where  $\kappa=1$ and $c=1.$
For chosen gauge
formulas
(\ref{riccg})-(\ref{scalg}) take the form:
\begin{equation}\label{curvc}
R^0_0=\frac{3}{a^2}\left(\frac{\dot a^2}{a^2}-\frac{\ddot
a}{a}\right);\quad
R^1_1=R^2_2=R^3_3=-\frac{1}{a^2}\left(\frac{\dot a^2}{a^2}+\frac{\ddot
a}{a}+2k^2\right);\quad R=-\frac{6}{a^3}\left(\ddot a+k^2
a\right),
\end{equation}
and Einstein equations (\ref{eq10}) are expressed by the formulas:
\begin{equation}\label{eq20}
\frac{3}{a^2}\left(\frac{\dot a^2}{a^2}+k^2\right)=\varepsilon+\Lambda;\quad
\frac{1}{a^2}\left(2\frac{\ddot a}{a}-\frac{\dot
a^2}{a^2}+k^2\right)=-p+\Lambda.
\end{equation}
With using (\ref{eq20}) it is easily to check validity of the relation:
\begin{equation}\label{cons}
\dot\varepsilon=-3(p+\varepsilon)\frac{\dot a}{a},
\end{equation}
having sense of energy conservation in adiabatically evolving Universe.
For equation of state (\ref{eqstat}) relation  (\ref{cons})
leads to the integral:
\begin{equation}\label{int0}
\varepsilon=ca^{-3(1+\alpha)},
\end{equation}
where $c$ --- integration constant. Combining (\ref{int0}) and
first equation in (\ref{eq20}), we go to the useful formula:
\begin{equation}\label{f1}
\dot a=a\sqrt{\frac{ca^{-1-3\alpha}}{3}+\frac{\Lambda a^2}{3}-k^2}.
\end{equation}
Differentiating this formula and expressing derivative $\dot a$
again through this formula, we go to another useful formula:
\begin{equation}\label{f2}
\ddot a=\frac{c(1-3\alpha)}{6}a^{-3\alpha}-k^2a+\frac{2}{3}\Lambda a^3.
\end{equation}
By the same manner one can obtain the following useful formulas, which will be used
in next sections:
\begin{equation}\label{ff1}
R=(3\alpha-1)ca^{-3(1+\alpha)}-4\Lambda;\quad
R_0^0=\frac{c(1+3\alpha)}{2}a^{-3(1+\alpha)}-\Lambda;\quad
R_i^i=\frac{c(\alpha-1)}{2}a^{-3(1+\alpha)}-\Lambda;
\end{equation}
\begin{equation}\label{ff2}
\dot
R=-3c(1+\alpha)(3\alpha-1)a^{-3(1+\alpha)}\sqrt{\frac{ca^{-1-3\alpha}}{3}+\frac{\Lambda a^2}{3}-k^2};
\end{equation}
\begin{equation}\label{ff3}
\ddot
R=-3c(1+\alpha)(3\alpha-1)a^{-3(1+\alpha)}(3(1+\alpha)k^2-\frac{c(7+9\alpha)}{6}a^{-1-3\alpha}-\frac{1}{3}(3\alpha+2)\Lambda).
\end{equation}

Equation (\ref{f1}) under $\Lambda=0$ can be integrated in elementary
functions\footnote{Though integral for world time:
$\tau=\int_0^t a(t)\,dt$ can be expressed in elementary functions
only for some particular values of parameters $\alpha$ and $k.$}.
The result has the form:
\begin{equation}\label{solg}
a(t)=\left\{
\begin{array}{lcr}
(\sqrt{c/3}\sin[k(1+3\alpha)t/2]/k)^{2/(1+3\alpha)},&\text{при}&\alpha\neq-1/3;\\
& & \\
a_0\exp[\pm\sqrt{c/3-k^2}t],&\text{при}&\alpha=-1/3.
\end{array}\right.
\end{equation}
Simple analysis of the integral (\ref{f1}) shows,
that in physical region $(\alpha\ge-1)$
$\Lambda-$term  plays no role in vicinity of cosmological singularity, but it
dominates under $a\to\infty$ for flat and open cosmological models.

At the end of this section we perform  nonzero
components of second covariant derivatives (acting on scalar
functions)
$\nabla_{\alpha\beta}\equiv\nabla_\alpha\nabla_\beta$ in FRW-models:
\begin{equation}\label{formulas}
\nabla_{00}=\frac{d^2}{dt^2}-\frac{\dot
a}{a}\frac{d}{dt};\quad\nabla_{ik}=g_{ik}\frac{\dot
a}{a^3}\frac{d}{dt};\quad \Box\equiv\frac{1}{\sqrt{-g}}\frac{\partial}{\partial
x^\alpha}\sqrt{-g}g^{\alpha\beta}\frac{\partial}{\partial
x^\beta}=\frac{1}{a^2}\left(\frac{d^2}{dt^2}+\frac{2\dot
a}{a}\frac{d}{dt}\right).
\end{equation}

\section{Generalized cosmological models}\label{non}

In present paper we consider generalized cosmological models
of the following two kinds:
\begin{enumerate}
\item
Models with nonlinear equation of state;
\item
Models with nonlinear on curvature action for gravity.
\end{enumerate}

\subsection{Nonlinear equation of state}

Lets consider barotropic  fluid with equation of state of the following kind:
\begin{equation}\label{nonl}
p=\alpha\varepsilon(1+\sigma\varepsilon^b),
\end{equation}
where $\sigma$ and $b$ are parameters, responsible for nonlinear properties
of the matter.
Under $\sigma\to0$  equation (\ref{nonl})
goes\footnote{For the case of dust-like matter $\alpha=0$
it is necessary to use the following limit procedure in (\ref{nonl}): $\alpha\to0,$
$\sigma\to\infty,$ $\alpha\sigma<\infty.$}
to the standard equation (\ref{eqstat}). Physical nature of
the nonlinearity can be concerned with the more complicated mechanisms of interaction
of matter particles, quantum effects, unknown fundamental interactions,
etc.

If energy density of such nonlinear matter  has structural perturbation  $\delta\varepsilon,$
then its pressure will have structural perturbation of the kind:
\begin{equation}\label{dist1}
\delta p=\alpha\delta\varepsilon+\alpha\sigma\varepsilon^{b+1}.
\end{equation}
In this expression nonlinear term is not varied, since we
assume, that it already has first order of smallness
(with respect to $\sigma$).

\subsection{Nonlinear models of gravity}

Lets consider general scheme of nonlinear theories of gravity in cosmological context.
For space-time of FRW-models there is the only scalar invariant ---
scalar curvature $R.$
So, general kind of action for nonlinear cosmological models,
generalizing Einstein-Hilbert action, can be expressed by the formula:
\begin{equation}\label{ng}
S_{\text{g}}=-\frac{1}{2}\int\ast f(R),
\end{equation}
where $f(R)$ --- arbitrary
function of scalar curvature. Standard variational
procedure\footnote{Under varying Lagrange density
$f(R)\sqrt{-g}$ it is helpful to use Palatini's identity:
$\delta
R_{\alpha\beta}=\nabla_\gamma\delta\Gamma^\gamma_{\alpha\beta}-\nabla_\beta\delta\Gamma^\gamma_{\alpha\gamma}.$}
leads to the following expression for variational derivative of
(\ref{ng}) with respect to metric:
\begin{equation}\label{neeq}
\frac{1}{\sqrt{-g}}\frac{\delta S_{\text{g}}}{\delta
g^{\alpha\beta}}\equiv-\frac{1}{2}\mathcal{G}_{\alpha\beta}=-\frac{1}{2}([R_{\alpha\beta}-\nabla_\alpha\nabla_\beta+g_{\alpha\beta}\Box]f'(R)-\frac{1}{2}g_{\alpha\beta}f(R)).
\end{equation}
where $\mathcal{G}$ is generalized Einstein tensor.
Let  $f(R)$ can be expressed through the generalized Taylor decomposition:
\begin{equation}\label{teyl}
f(R)=2\Lambda+R+\sum\limits_{s\neq0,1}\lambda_s R^s,
\end{equation}
where $2\Lambda=\lambda_0,$ $\lambda_1=1$ and $s$
--- any real number, excepting 0 and 1.
Choice of the $\lambda_1$ corresponds to our main problem:
to investigate structural stability of cosmological models in vicinity of standard FRW-models.
Choice of the $\lambda_0$ will allow us to investigate
structural stability of $\Lambda$-term cosmology  in frame of general scheme.
So, in space of nonlinear theories of gravity we are interested by those theories,
which are characterized by the following infinitely dimensional
vector of parameters:
$(\dots\lambda_{s_{-1}},\lambda_0,1,\lambda_{s_2},\dots),$ when
\footnote{Note, that all $\lambda_s$ have different dimensions:
$[\lambda_s]=\ell^{2(s-1)}.$
In spite of conditional sense of "smallness"\, for any dimensional value,
our analysis in vicinity of the point $(\dots,0,1,0,\dots)$   in parametric space
is absolutely correct.}
all $\lambda_{s_i}\to0$ under $i\neq0,1.$
In view of  this aim it is expediently
to decompose (with using (\ref{teyl})) tensor $\mathcal{G}$
on standard Einstein tensor
 (\ref{eing}) and
"perturbing"\, part, which can be interpreted (up to a constant) as
energy-momentum tensor $T^{\text{n}}$ of nonlinear self-action:
\begin{equation}\label{rep1}
\mathcal{G}_{\alpha\beta}=G_{\alpha\beta}+\Delta G_{\alpha\beta}=G_{\alpha\beta}+
[\stackrel{\circ}{R}_{\alpha\beta}-
\stackrel{\circ}{\nabla}_\alpha\stackrel{\circ}{\nabla}_\beta+
\stackrel{\circ}{g}_{\alpha\beta}\stackrel{\circ}{\Box}]F'(\stackrel{\circ}{R})
-\frac{1}{2}\stackrel{\circ}{g}_{\alpha\beta}F(\stackrel{\circ}{R}),
\end{equation}
where
\[
F(\stackrel{\circ}{R})=2\Lambda+\sum\limits_{s\neq0,1}\lambda_s {\stackrel{\circ}{R^s}},
\]
and circle over values relates them to unperturbed cosmological metric
of those FRW-model, whose structural stability we are studying.

Typical $s$-th
member of this nonlinear perturbation in right-hand side of perturbed
Einstein equation will have the following kind:
\begin{equation}\label{distn}
-(\Delta G_s)^\alpha_\beta=(T^{\text{n}}_s)^\alpha_\beta=-s\lambda_s(\stackrel{\circ}{R^\alpha_\beta}-\stackrel{\!\circ}{\nabla^\alpha}\stackrel{\!\circ}{\nabla_\beta}+
\delta^\alpha_\beta\stackrel{\circ}{\Box})\stackrel{\!\!\circ}{R^{s-1}}
+\frac{\lambda_s}{2}\delta^\alpha_\beta{\stackrel{\circ}{R^s}}.
\end{equation}
With using (\ref{curvc}), (\ref{int0})-(\ref{ff3}) and
(\ref{formulas}) after some algebra, we go to the following expressions
for non-zero components of nonlinear self-action tensor  (there is no summation over repeating indexes!):
\begin{equation}\label{dist3}
(T^{\text{n}}_s)^\beta_\beta=\lambda_s(3\alpha-1)^{s-1}c^sa^{-3s(1+\alpha)}(A_s^\beta+B_s^{\beta}a^{1+3\alpha}),
\end{equation}
where
\begin{equation}\label{coeff1}
A_s^0=\frac{1}{2}(3\alpha(2s^2-3s+1)+6s^2-7s-1);\quad
B_s^0=-\frac{9k^2}{c}s(s-1)(1+\alpha);
\end{equation}
\begin{equation}\label{coeff2}
A_s^1=A_s^2=A_s^3=-(s+s\alpha-1)A_s^0;\quad
B_s^1=B_s^2=B_s^3=-(s+s\alpha-\alpha-\frac{4}{3})B_s^0
\end{equation}
Note, that all formulas are valid for any real value of parameter
$s.$

\section{Equations for structural perturbations} \label{ge}

In order to derive linear differential equation for structural
perturbations it is necessary  to extract null order
in the left-hand (geometrical) side  of structurally perturbed Einstein equations:
\begin{equation}\label{stein}
\hat G(a(t))=T^{\text{n}}+T^{\text{mat}},
\end{equation}
where we consider now Einstein tensor as nonlinear differential operator
$\hat G,$
acting on space of scale factors.
Introducing structural perturbation of scale factor: $a(t)\to
a(t)+\delta(t),$ we go to the formula:
\begin{equation}\label{disein}
\hat G(a(t)+\delta(t))=\hat G(a(t))+\delta \hat
G(\delta(t),a(t))+o(\delta),
\end{equation}
where
\begin{equation}\label{dG}
\delta \hat
G(\delta(t),a(t))\equiv\frac{d}{d\varepsilon}\hat
G(a(t)+\epsilon\delta(t))|_{\epsilon=0}
\end{equation}
--- "differential of Einstein operator", which is
linear differential operator on its first argument.
Direct calculations with using formulas
(\ref{eq20}), (\ref{dG}) and
(\ref{int0})-(\ref{f2}) lead to the following nonzero
components of
$\delta G$:
\begin{equation}\label{dGc}
\delta
G_0^0(\delta,a)=\frac{6}{a^3}\left(\sqrt{ca^{-1-3\alpha}/3-k^2}\dot\delta+(k^2-2ca^{-1-3\alpha}/3)\delta\right);
\end{equation}
\begin{equation}\label{dGc1}
\delta
G_1^1(\delta,a)=\frac{2}{a^3}\ddot\delta-\frac{2}{a^3}\sqrt{ca^{-1-3\alpha}/3-k^2}\dot\delta+\frac{c(1+9\alpha)}{3}a^{-4-3\alpha}\delta.
\end{equation}
Substituting it into equations (\ref{stein}) and keeping in mind, that
in null order on $\delta$ these equations
lead to Einstein equations
(\ref{eq20}) for unperturbed FRW-model  which are satisfied identically,
we go to the following linearized system of equations on structural
perturbations:
\begin{equation}\label{structg1}
\frac{6}{a^3}\left(\sqrt{ca^{-1-3\alpha}/3-k^2}\dot\delta+(k^2-2ca^{-1-3\alpha}/3)\delta\right)
=(T^{\text{n}})_0^0+\delta\varepsilon;
\end{equation}
\begin{equation}\label{structg2}
\frac{2}{a^3}\ddot\delta-\frac{2}{a^3}\sqrt{ca^{-1-3\alpha}/3-k^2}\dot\delta+
\frac{c(1+9\alpha)}{3}a^{-4-3\alpha}\delta=(T^{\text{n}})_1^1-\delta p.
\end{equation}
Using formula (\ref{dist1}) and excluding from
(\ref{structg1})-(\ref{structg2}) structural perturbation $\delta\varepsilon,$
we obtain unique basic equation:
\begin{equation}\label{ggen}
2\ddot\delta+2(3\alpha-1)\sqrt{ca^{-1-3\alpha}/3-k^2}\dot\delta+(6\alpha
k^2+c(1-3\alpha)a^{-1-3\alpha}/3)\delta=
\end{equation}
\[
a^3(\alpha(T^{\text{n}})_0^0+(T^{\text{n}})_1^1-\alpha\sigma\varepsilon^{b+1}).
\]
Going to the new variable
$x=a(t),$
equation (\ref{ggen}) can be transformed to the form:
\begin{equation}\label{ggen1}
2x^2(cx^{-1-3\alpha}/3-k^2)\delta''+(\delta-x\delta')(6\alpha
k^2+c(1-3\alpha)x^{-1-3\alpha}/3)=
\end{equation}
\[
x^3(\alpha(T^{\text{n}})_0^0+(T^{\text{n}})_1^1)-\alpha\sigma c^{b+1}x^{-3(\alpha+1)(b+1)+3}.
\]

Since sources of structural perturbations in first order act independently from each
other, we can analyze structural stability of FRW-models by means of equation
(\ref{ggen1}) separately for every source.
Limit correspondence considerations suggest that structural perturbations
of scale factor must be determined by those particular
solutions of differential equation (\ref{ggen1}) which vanish
when source of perturbations vanishes\footnote{So,
we omit general solutions to homogeneous equation (\ref{ggen1}) (without its right-hand side).
It is interesting to note, that this solutions
define class of infinitesimal
{\it fantom} conformal perturbations,  which up to high orders of smallness
transform  apparently different metrics, originated from  the same
matter source, to each other.}.

\subsection{Structural stability of cosmological models within nonlinear gravity}

Let consider structural perturbations of $f(R)$-cosmological models.
Perturbation from $s$-th member of Taylor decomposition
in our notations takes the form:
\begin{equation}\label{disturb1}
x^3(\alpha((T^{\text{n}}_s)_0^0+(T^{\text{n}}_s)_1^1)=\lambda_s(3\alpha-1)^{s-1}c^sx^{-3s(1+\alpha)+3}(A_s+B_sx^{1+3\alpha}),
\end{equation}
where
\[
A_s=\alpha A^0_s+A^1_s=(\alpha+1)(1-s)A^0_s;\quad B_s=\alpha
B^0_s+B^1_s=(2\alpha+\frac{4}{3}-s-s\alpha)B^0_s.
\]
Looking at equation (\ref{ggen1}) with right-hand side (\ref{disturb1})
together with  (\ref{coeff1})-(\ref{coeff2}),
one can immediately conclude:
{\it all standard cosmological models with isotropic radiation $(\alpha=1/3)$
and vacuum-like matter $(\alpha=-1)$ are structurally stable
in any order on curvature,} since this concrete values of $\alpha$
imply vanishing of sources of structural perturbations.

In general case let introduce the following notations:
\[
\xi(x)=\frac{\delta(x)}{3\lambda_s(3\alpha-1)^{s-1}c^{s-1}(1+\alpha)(1-s)x};\quad
\bar k^2=\frac{3k^2}{c}.
\]
General condition of structural stability is {\it boundedness  of relative perturbation
$\xi(x)$ for any allowed   $x.$}
Note, that in this section we accept, that  $\alpha\neq-1$ and
$\alpha\neq1/3$ in view of above mentioned specific properties
of FRW-models with such parameters. Also we put anywhere\footnote{Situation with
$s=1$ has no particular interest, since, in fact,
it can be  reduced to redefinition of Einstein gravitational constant
and so it can not influence on stability.} $s\neq1.$

In order to analyze  $S$-stability  let consider
asymptotic kind of equation
(\ref{ggen1}) together with its right-hand side (\ref{disturb1}) under $x\to0$ and under $\alpha>-1/3,$
when cosmological singularity does take place
(solution (\ref{solg})).
For general values of parameters this asymptotic kind
is expressed by the following equation:
\begin{equation}\label{sold1}
2x^2(x\xi(x))''+(1-3\alpha)x(\xi(x)-(x\xi(x))')=
\frac{1}{2}(3\alpha(2s^2-3s+1)+6s^2-7s-1)x^{-3s(1+\alpha)+4+3\alpha}.
\end{equation}
Its particular solution, which we are interested, has the form:
\begin{equation}\label{term1}
\xi(x)|_{x\to0}=
\frac{6\alpha s^2-9s\alpha+3\alpha+6s^2-7s-1}{6(6s+6s\alpha-7-9\alpha)(1+\alpha)(s-1)}
x^{-3(s-1)(1+\alpha)}.
\end{equation}
So, condition of  $S$-stability is positiveness
of index: $-3(s-1)(1+\alpha)>0,$ that under
$1+3\alpha>0$ takes place, when  $s<1.$ In other words, {\it almost all
nonlinear analytic generalizations of standard cosmology
possess  structural instability of singularity.}
Note, that non-analytic nonlinear models (for example, containing in lagrangians
negative powers
of  $R$) may be  $S$-stable.

The words
"almost all"\, are necessary to separate $\Lambda$-term cosmology $(s=0)$
and some particular cases, which are $S$-stable.
$S$-stability of the particular case,
when coefficient in (\ref{term1})
vanishes:
\begin{equation}\label{part1}
6\alpha s^2-9s\alpha+3\alpha+6s^2-7s-1=0,
\end{equation}
corresponds to the asymptotic equation (\ref{sold1}) with other right-hand side:
\begin{equation}\label{sold2}
2x^2(x\xi(x))''+(1-3\alpha)x(\xi(x)-(x\xi(x))')=
3\bar k^2s(2\alpha+\frac{4}{3}-s(1+\alpha))x^{-3s(1+\alpha)+5+6\alpha}.
\end{equation}
Using relation between parameters:
\begin{equation}\label{rell}
\alpha=-\frac{6s^2-7s-1}{3(2s^2-3s+1)},
\end{equation}
which follows from  (\ref{part1}),
we obtain the following particular solution to equation (\ref{sold2}):
\begin{equation}\label{term2}
\xi(x)|_{x\to0}=-\frac{\bar{k}^2s(2s-1)(s-1)}{2(4s^2-7)}x^{-2(s^2+s-3)/(2s-1)(s-1)}.
\end{equation}
Omitting the situations with $s<1,$ when almost all models
are $S$-stable, we obtain
 two new classes of $S$-stable models. The first one
is obtained under
$s\in(1;s_\ast],$ that corresponds to monotonic variation of
$\alpha\in(\infty;\alpha_\ast].$
Values of boundaries of the intervals are:
\begin{equation}\label{bounce}
s_\ast=\frac{-1+\sqrt{13}}{2}\approx1.30;\quad
\alpha_\ast=-\frac{47-13\sqrt{13}}{3(19-5\sqrt{13})}
\approx-0.04.
\end{equation}
Second class is obtained for flat models $\bar k=0.$
The interval of nontrivial  $S$-stability (i.e., interval, where
$\alpha>-1/3$) is described by the formula
(\ref{rell}) for $s\in(1;(1+\sqrt3)/2).$

Finally, the last particular case corresponds to the following relation between parameters:
\begin{equation}\label{part2}
(6s+6s\alpha-7-9\alpha)=0\Leftrightarrow\alpha=-\frac{6s-7}{3(2s-3)}\Leftrightarrow
s=\frac{7+9\alpha}{6(1+\alpha)},
\end{equation}
when denominator in (\ref{term1}) vanishes.
Asymptotic particular solution in this case has the form:
\begin{equation}\label{term3}
\xi(x)|_{x\to0}=\frac{(1-3\alpha)((1+3\alpha)\ln(x)+2)}{2(1+3\alpha)^2}x^{-(1+3\alpha)/2}.
\end{equation}
Under $\alpha>-1/3$ $(1<s<3/2)$ power is negative and the models are $S$-unstable, while
under $s<1$ there is no singularity and the models are  $S$-stable.

\begin{figure}[htb]
\footnotesize
\centering \unitlength=0.50mm \special{em:linewidth 0.4pt}
\linethickness{0.4pt}
\unitlength=1mm
\special{em:linewidth 0.4pt}
\linethickness{0.4pt}
\begin{picture}(78.00,59.33)
\emline{0.83}{5.00}{1}{73.83}{5.00}{2}
\emline{73.83}{5.00}{3}{69.33}{5.83}{4}
\emline{69.33}{4.17}{5}{74.00}{5.00}{6}
\emline{14.50}{14.83}{7}{14.50}{54.17}{8}
\emline{14.50}{13.83}{9}{14.50}{15.00}{10}
\emline{14.50}{13.83}{11}{66.17}{13.83}{12}
\emline{15.50}{15.00}{13}{54.50}{15.00}{14}
\emline{55.50}{15.00}{15}{66.00}{15.00}{16}
\bezier{160}(15.50,22.50)(33.50,17.67)(54.50,17.17)
\bezier{160}(15.50,23.67)(34.00,18.67)(54.50,18.17)
\emline{15.50}{15.00}{17}{15.50}{22.50}{18}
\emline{15.50}{23.83}{19}{15.50}{54.33}{20}
\emline{54.67}{15.00}{21}{54.67}{17.17}{22}
\emline{54.67}{18.17}{23}{54.67}{54.17}{24}
\emline{55.67}{15.00}{25}{55.67}{17.17}{26}
\emline{55.67}{18.33}{27}{55.67}{54.17}{28}
\emline{15.00}{13.00}{29}{15.00}{11.00}{30}
\emline{15.00}{10.00}{31}{15.00}{8.00}{32}
\emline{15.00}{7.00}{33}{15.00}{5.00}{34}
\emline{55.00}{13.00}{35}{55.00}{11.00}{36}
\emline{55.00}{10.00}{37}{55.00}{8.00}{38}
\emline{55.00}{7.00}{39}{55.00}{5.00}{40}
\emline{27.00}{19.00}{41}{27.00}{17.00}{42}
\emline{27.00}{13.00}{43}{27.00}{11.00}{44}
\emline{27.00}{10.00}{45}{27.00}{8.00}{46}
\emline{27.00}{7.00}{47}{27.00}{5.00}{48}
\emline{28.00}{21.00}{49}{30.00}{21.00}{50}
\emline{31.00}{21.00}{51}{33.00}{21.00}{52}
\emline{16.00}{24.00}{53}{18.00}{24.00}{54}
\emline{19.00}{24.00}{55}{21.00}{24.00}{56}
\emline{22.00}{24.00}{57}{24.00}{24.00}{58}
\emline{25.00}{24.00}{59}{27.00}{24.00}{60}
\emline{28.00}{24.00}{61}{30.00}{24.00}{62}
\emline{31.00}{24.00}{63}{33.00}{24.00}{64}
\emline{14.50}{53.17}{65}{15.50}{53.17}{66}
\emline{14.50}{51.00}{67}{15.50}{51.00}{68}
\emline{14.50}{49.00}{69}{15.50}{49.00}{70}
\emline{14.50}{46.83}{71}{15.50}{46.83}{72}
\emline{14.50}{45.00}{73}{15.50}{45.00}{74}
\emline{14.50}{43.00}{75}{15.50}{43.00}{76}
\emline{14.50}{41.00}{77}{15.50}{41.00}{78}
\emline{14.50}{39.00}{79}{15.50}{39.00}{80}
\emline{14.50}{37.00}{81}{15.50}{37.00}{82}
\emline{14.50}{35.00}{83}{15.50}{35.00}{84}
\emline{15.50}{35.00}{85}{15.50}{35.00}{86}
\emline{14.50}{33.00}{87}{15.50}{33.00}{88}
\emline{14.50}{31.00}{89}{15.50}{31.00}{90}
\emline{14.50}{29.00}{91}{15.50}{29.00}{92}
\emline{14.50}{27.00}{93}{15.50}{27.00}{94}
\emline{14.50}{25.00}{95}{15.50}{25.00}{96}
\emline{14.50}{23.00}{97}{15.50}{23.00}{98}
\emline{14.50}{21.00}{99}{15.50}{21.00}{100}
\emline{14.50}{19.00}{101}{15.50}{19.00}{102}
\emline{14.50}{17.00}{103}{15.67}{17.00}{104}
\emline{14.50}{15.00}{105}{15.50}{15.00}{106}
\emline{15.50}{23.83}{107}{15.50}{22.67}{108}
\emline{17.00}{23.33}{109}{17.00}{22.17}{110}
\emline{19.00}{22.83}{111}{19.00}{21.67}{112}
\emline{21.00}{22.33}{113}{21.00}{21.17}{114}
\emline{23.00}{21.67}{115}{23.00}{20.50}{116}
\emline{25.00}{21.33}{117}{25.00}{20.17}{118}
\emline{27.00}{21.00}{119}{27.00}{19.83}{120}
\emline{16.00}{52.00}{121}{18.00}{54.00}{122}
\emline{16.00}{46.00}{123}{24.00}{54.00}{124}
\emline{16.00}{40.00}{125}{30.00}{54.00}{126}
\emline{16.00}{34.00}{127}{36.00}{54.00}{128}
\emline{16.00}{28.00}{129}{42.00}{54.00}{130}
\emline{18.00}{24.00}{131}{48.00}{54.00}{132}
\emline{22.00}{22.00}{133}{54.00}{54.00}{134}
\emline{28.00}{22.00}{135}{54.00}{48.00}{136}
\emline{32.00}{20.00}{137}{54.00}{42.00}{138}
\emline{38.00}{20.00}{139}{54.00}{36.00}{140}
\emline{44.00}{20.00}{141}{54.00}{30.00}{142}
\emline{50.00}{20.00}{143}{54.00}{24.00}{144}
\emline{16.00}{16.00}{145}{20.00}{20.00}{146}
\emline{22.00}{16.00}{147}{26.00}{20.00}{148}
\emline{28.00}{16.00}{149}{30.00}{18.00}{150}
\emline{34.00}{16.00}{151}{36.00}{18.00}{152}
\emline{40.00}{16.00}{153}{42.00}{18.00}{154}
\put(15.00,4.33){\makebox(0,0)[ct]{$-1/3$}}
\put(55.00,4.33){\makebox(0,0)[ct]{$1/3$}}
\put(27.00,4.50){\makebox(0,0)[ct]{$\alpha_\ast$}}
\put(71.17,6.33){\makebox(0,0)[lb]{$\alpha$}}
\put(36.50,57.33){\makebox(0,0)[lc]{$s$}}
\put(35.67,20.50){\makebox(0,0)[lb]{$s_\ast$}}
\put(35.50,25.33){\makebox(0,0)[lb]{$s_{\ast\ast}$}}
\put(35.83,12.83){\makebox(0,0)[lt]{$1$}}
\put(21.17,20.00){\makebox(0,0)[ct]{$\bar k^2=0$}}
\put(13.83,19.67){\makebox(0,0)[rc]{$\bar k^2<0$}}
\put(13.83,38.33){\makebox(0,0)[rc]{$\bar k^2>0$}}
\emline{56.00}{50.00}{155}{60.00}{54.00}{156}
\emline{56.00}{44.00}{157}{66.00}{54.00}{158}
\emline{56.00}{38.00}{159}{72.00}{54.00}{160}
\emline{56.00}{32.00}{161}{78.00}{54.00}{162}
\emline{56.00}{26.00}{163}{76.00}{46.00}{164}
\emline{56.00}{20.00}{165}{76.00}{40.00}{166}
\emline{60.00}{18.00}{167}{76.00}{34.00}{168}
\emline{66.00}{18.00}{169}{76.00}{28.00}{170}
\emline{72.00}{18.00}{171}{76.00}{22.00}{172}
\emline{66.00}{15.00}{173}{76.00}{15.00}{174}
\emline{66.33}{13.83}{175}{75.83}{13.83}{176}
\bezier{76}(55.67,18.17)(64.17,17.17)(75.00,17.50)
\bezier{80}(55.67,17.17)(62.83,16.50)(75.17,16.50)
\emline{46.67}{17.00}{177}{45.33}{15.67}{178}
\emline{53.33}{16.83}{179}{52.00}{15.50}{180}
\emline{58.67}{16.50}{181}{57.67}{15.50}{182}
\emline{35.00}{1.67}{183}{35.00}{59.33}{184}
\emline{35.00}{59.33}{185}{34.17}{55.17}{186}
\emline{35.83}{55.17}{187}{35.00}{59.33}{188}
\put(34.17,4.17){\makebox(0,0)[rt]{$O$}}
\end{picture}
\caption{\small Diagram of $S$-stability of FRW-models. Inclined dash
points region of  $S$-instability of FRW-models. Boundary of this region
is double  line. At this boundary FRW-models can be both $S$-stable
(vertical line $\alpha=1/3,$ horizontal line  $s=1$ and part of rational dependency
(\ref{rell}), beginning at the point $\{s_\ast;\alpha_\ast\}$ (formula (\ref{bounce}))
and conditionally  $S$-stable (vertical line (with horizontal dash) $\alpha=-1/3$ with condition
(\ref{part4}); upper part of this line correspondes to closed models, lower --- to the open ones;
and part (with vertical dash) of dependency  (\ref{rell}),
lying between  $s_\ast$ and $s_{\ast\ast}=(1+\sqrt{3})/2$ with condition $\bar k^2=0$).}\label{d1}
\end{figure}
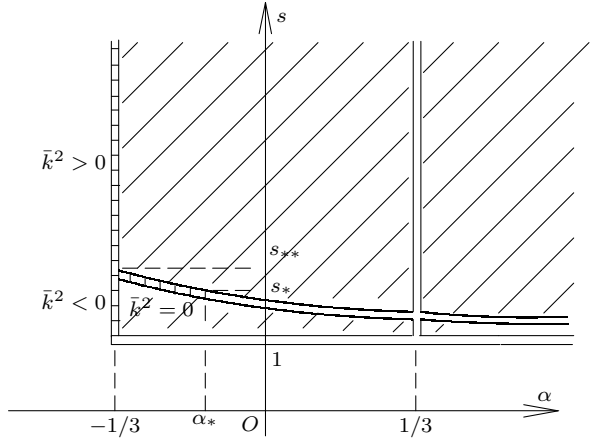

So far the case $\alpha=-1/3$ has remained out of our consideration.
Particular solution
to the equation
(\ref{ggen1}) with right-hand side  (\ref{disturb1}), taken under $\alpha=-1/3,$
has the form:
\begin{equation}\label{term4}
\xi(x)=\frac{-2s^2+2s+1-2{\bar k}^2s+2{\bar k}^2s^2}{8({\bar k}^2-1)(s-1)^2}x^{-2s+2},
\end{equation}
that leads to the conclusion: {\it under $s>1$ generalized nonlinear models with equation
of state: $p=-\varepsilon/3$ are $S$-unstable and $A$-stable}
and vise verse: {\it under $s<1$ such models are $S$-stable and $A$-unstable.}
Particular situation arises when coefficient in expression (\ref{term4}) vanishes.
In this case the model is structurally stable in all senses, at the least, in linear approximation.
In closed models this situation is described by relation:
\begin{equation}\label{part4}
s=\frac{{\bar k}^2-1\pm\sqrt{{\bar k}^4-4{\bar k}^2+3}}{2({\bar k}^2-1)}.
\end{equation}
Since solution (\ref{solg}) must be real, parameter ${\bar k}$
lies in interval $(0;1).$  Parameter $s$ (on both branches of square
root)
belongs to the union  $(-\infty;(1-\sqrt3)/2]\cup[(1+\sqrt3)/2;\infty).$
For open models
relation between $s$ and ${\bar k}$ can be obtained
from (\ref{part4}) by the substitution  ${\bar k}^2\to -{\bar k}^2.$
When  ${\bar k}$ varies from  $0$ to $\infty$ parameter
$s$ monotonically falls from $(1+\sqrt3)/2$ to $1$ on upper branch of square root
and monotonically increases from
$(1-\sqrt3)/2$ to $0$ on lower one.
The value $\bar k^2=1$ is singular for  equation (\ref{ggen1}),
since under $\alpha=-1/3$ and $\bar k^2=1$ this equation become algebraic: $x^{-2s+3}=0.$
It corresponds to the limit $s\to\infty,$ which
is reached on the upper branch of solution (\ref{part4}).
In this degenerate case cosmological models can be
both structurally stable, and structurally unstable,
since basic equation (\ref{ggen1}) is satisfied by any function $\delta(x).$
All results are summarized in the diagram of  $S$-stability (fig.\ref{d1}).

In order to analyze  $A$-stability of  FRW-models
it is necessary to go to asymptotic kind of equation (\ref{ggen1}) and of its right-hand side
(\ref{disturb1}) under $x\to\infty.$ Note, that
energetic condition: $1+3\alpha\ge0$ is unnecessary now.

In first, lets note, that under $1+3\alpha<0$
asymptotic kind of equation for structural perturbations
exactly coincides  with (\ref{sold1}), while its particular solution coincides with (\ref{term1}),
which one should consider now as solution under $x\to\infty.$
Analysis of power sign leads to the conclusion, that {\it under $-1<\alpha<-1/3$ almost
all FRW-models are $A$-stable  when $s>1$ and  are $A$-instable when $s<1$, and under
$\alpha<-1$ vise verse:
almost all FRW-models are $A$-instable when  $s>1$ and $A$-stable when $s<1$.}
Lets go to the particular situations.
The case, when coefficient in (\ref{term1})
vanishes is described by the formulas (\ref{part1})-(\ref{term2}),
which should be considered as asymptotic expressions under $x\to\infty$
and $\alpha<-1/3.$ Additional  $A$-stable branch of FRW-models arises here
when $s\in(-\infty;s^\ast]\cup(2;\infty),$
where $s^\ast=-(1+\sqrt{13})/2\approx-2.30.$ Here
on the first interval  $\alpha\in(-1;\alpha^\ast],$
where $\alpha^\ast=-(4+\sqrt{13})/9\approx-0.85$ and $\alpha$ monotonically
increases.
On the second interval $\alpha$ decreases from  $-1$ to its minimal value
$\alpha_{\text{min}}=(4\sqrt6-13)/3\approx-1.07,$
which is reached under $s_{\text{min}}=2+\sqrt6/2\approx3.22,$ and then
monotonically increases to the value $-1.$
There is no new $A$-stable models in case of vanishing of denominator in
 (\ref{term1}).

Now lets go to the case $1+3\alpha>0.$
Asymptotic kind of equation for structural perturbations
under $\bar k^2\neq0$ takes the form:
\begin{equation}\label{eqq}
-2x^2(x\xi(x))''+6\alpha x(\xi(x)-(x\xi(x))')=
3s(2\alpha+\frac{4}{3}-s(1+\alpha))x^{-3s(1+\alpha)+4+3\alpha}.
\end{equation}
Its particular solution:
\begin{equation}\label{solqq}
\xi(x)|_{x\to\infty}=\frac{s x^{-3(s-1)(1+\alpha)}}{6(s-1)(1+\alpha)}.
\end{equation}
So,  {\it under $\alpha>-1/3$ almost all non-flat FRW-models
are $A$-stable under $s>1$ and are $A$-unstable under $s<1.$} In particular case
$s=0$ equation  (\ref{eqq}) will have another right-hand side and another particular solution:
\[
\xi(x)|_{x\to\infty}=-\frac{3\alpha-1}{24q^2(1+\alpha)}x^2,
\]
that implies $A$-instability of all such models.

Finally, for the case of flat models equation for structural
perturbation become identical to (\ref{sold1}) and its particular
solution become identical to  (\ref{term1}). So, the models are $A$-stable under
$s>1.$ The relation (\ref{rell}) defines
additional branch of $A$-stable models in the region $\alpha>-1/3$ under $s^{\ast\ast}<s<1$,
where $s^{\ast\ast}=(1-\sqrt3)/2\approx-0.37,$ while relation
(\ref{part2}) defines additional branch of $A$-unstable models under $3/2>s>1.$

\begin{figure}[htb]
\footnotesize
\centering \unitlength=0.50mm \special{em:linewidth 0.4pt}
\linethickness{0.4pt}
\unitlength=1.00mm
\special{em:linewidth 0.4pt}
\linethickness{0.4pt}
\begin{picture}(79.67,80.00)
\emline{0.00}{40.00}{1}{79.67}{40.00}{2}
\emline{79.67}{40.00}{3}{75.83}{40.83}{4}
\emline{75.83}{39.00}{5}{79.67}{40.00}{6}
\emline{40.00}{0.00}{7}{40.00}{79.50}{8}
\emline{40.00}{79.50}{9}{39.00}{75.33}{10}
\emline{41.00}{75.33}{11}{40.00}{79.50}{12}
\emline{40.00}{55.00}{13}{41.17}{55.00}{14}
\emline{40.00}{70.00}{15}{41.17}{70.00}{16}
\emline{40.00}{25.00}{17}{41.17}{25.00}{18}
\emline{40.00}{10.00}{19}{41.17}{10.00}{20}
\emline{70.00}{40.00}{21}{70.00}{41.33}{22}
\emline{10.00}{40.00}{23}{10.00}{41.33}{24}
\emline{9.33}{0.00}{25}{9.33}{54.33}{26}
\emline{9.33}{55.67}{27}{9.33}{69.33}{28}
\emline{9.33}{70.67}{29}{9.33}{79.50}{30}
\emline{10.67}{0.00}{31}{10.67}{54.17}{32}
\emline{10.67}{55.67}{33}{10.67}{69.33}{34}
\emline{10.67}{70.50}{35}{10.67}{79.50}{36}
\emline{30.67}{0.00}{37}{30.67}{54.33}{38}
\emline{32.00}{0.00}{39}{32.00}{33.67}{40}
\emline{32.00}{35.50}{41}{32.00}{54.33}{42}
\emline{48.00}{54.33}{43}{48.00}{40.83}{44}
\emline{48.00}{39.33}{45}{48.00}{0.00}{46}
\emline{49.33}{0.00}{47}{49.33}{39.50}{48}
\emline{49.33}{41.00}{49}{49.33}{54.50}{50}
\emline{0.00}{54.33}{51}{9.33}{54.33}{52}
\emline{10.67}{54.33}{53}{30.67}{54.33}{54}
\emline{32.00}{54.33}{55}{48.00}{54.33}{56}
\emline{49.33}{54.33}{57}{79.67}{54.33}{58}
\emline{0.00}{55.83}{59}{9.33}{55.83}{60}
\emline{10.67}{55.83}{61}{30.67}{55.83}{62}
\emline{32.00}{55.83}{63}{79.67}{55.83}{64}
\bezier{36}(9.33,70.83)(6.83,74.83)(7.00,79.50)
\bezier{48}(9.50,69.00)(6.50,73.00)(5.67,79.50)
\emline{10.67}{70.83}{65}{10.67}{69.33}{66}
\bezier{28}(11.17,0.00)(11.17,3.67)(13.00,6.67)
\bezier{20}(12.33,0.67)(13.33,4.50)(14.17,5.67)
\emline{13.00}{6.67}{67}{14.33}{5.67}{68}
\bezier{68}(32.00,35.83)(36.50,38.33)(48.00,40.83)
\bezier{68}(32.00,33.50)(39.50,37.83)(48.00,39.17)
\bezier{104}(49.33,41.00)(58.83,43.50)(75.00,45.17)
\bezier{108}(49.33,39.33)(65.17,43.33)(75.33,43.83)
\bezier{64}(32.00,55.83)(38.50,57.50)(48.00,58.33)
\bezier{72}(30.67,55.83)(40.67,59.17)(48.00,59.50)
\bezier{112}(48.00,59.67)(73.17,62.33)(76.00,62.00)
\emline{41.00}{47.00}{69}{44.00}{47.00}{70}
\emline{45.00}{47.00}{71}{48.00}{47.00}{72}
\emline{50.00}{47.00}{73}{53.00}{47.00}{74}
\emline{54.00}{47.00}{75}{57.00}{47.00}{76}
\emline{58.00}{47.00}{77}{61.00}{47.00}{78}
\emline{62.00}{47.00}{79}{65.00}{47.00}{80}
\emline{66.00}{47.00}{81}{69.00}{47.00}{82}
\emline{70.00}{47.00}{83}{73.00}{47.00}{84}
\emline{74.00}{47.00}{85}{77.00}{47.00}{86}
\emline{41.00}{63.00}{87}{44.00}{63.00}{88}
\emline{45.00}{63.00}{89}{48.00}{63.00}{90}
\emline{49.00}{63.00}{91}{52.00}{63.00}{92}
\emline{53.00}{63.00}{93}{56.00}{63.00}{94}
\emline{57.00}{63.00}{95}{60.00}{63.00}{96}
\emline{61.00}{63.00}{97}{64.00}{63.00}{98}
\emline{65.00}{63.00}{99}{68.00}{63.00}{100}
\emline{69.00}{63.00}{101}{72.00}{63.00}{102}
\emline{73.00}{63.00}{103}{76.00}{63.00}{104}
\emline{33.00}{35.00}{105}{36.00}{35.00}{106}
\emline{37.00}{35.00}{107}{40.00}{35.00}{108}
\emline{14.00}{6.00}{109}{16.00}{6.00}{110}
\emline{17.00}{6.00}{111}{20.00}{6.00}{112}
\emline{21.00}{6.00}{113}{24.00}{6.00}{114}
\emline{25.00}{6.00}{115}{28.00}{6.00}{116}
\emline{29.00}{6.00}{117}{31.00}{6.00}{118}
\emline{14.00}{6.00}{119}{14.00}{9.00}{120}
\emline{14.00}{10.00}{121}{14.00}{13.00}{122}
\emline{14.00}{14.00}{123}{14.00}{17.00}{124}
\emline{14.00}{18.00}{125}{14.00}{21.00}{126}
\emline{14.00}{22.00}{127}{14.00}{25.00}{128}
\emline{14.00}{26.00}{129}{14.00}{29.00}{130}
\emline{14.00}{30.00}{131}{14.00}{33.00}{132}
\emline{14.00}{34.00}{133}{14.00}{37.00}{134}
\emline{14.00}{38.00}{135}{14.00}{40.00}{136}
\emline{11.00}{70.00}{137}{14.00}{70.00}{138}
\emline{15.00}{70.00}{139}{18.00}{70.00}{140}
\emline{19.00}{70.00}{141}{22.00}{70.00}{142}
\emline{23.00}{70.00}{143}{26.00}{70.00}{144}
\emline{27.00}{70.00}{145}{30.00}{70.00}{146}
\emline{31.00}{70.00}{147}{34.00}{70.00}{148}
\emline{35.00}{70.00}{149}{38.00}{70.00}{150}
\emline{39.00}{70.00}{151}{40.00}{70.00}{152}
\emline{6.00}{56.00}{153}{9.00}{62.00}{154}
\emline{2.00}{56.00}{155}{9.00}{69.00}{156}
\emline{0.00}{60.00}{157}{7.00}{72.00}{158}
\emline{0.00}{67.00}{159}{6.00}{76.00}{160}
\emline{0.00}{74.00}{161}{4.00}{80.00}{162}
\emline{7.00}{77.00}{163}{9.00}{80.00}{164}
\emline{8.00}{74.00}{165}{9.00}{76.00}{166}
\emline{11.00}{49.00}{167}{16.00}{54.00}{168}
\emline{11.00}{44.00}{169}{21.00}{54.00}{170}
\emline{11.00}{39.00}{171}{26.00}{54.00}{172}
\emline{11.00}{34.00}{173}{31.00}{54.00}{174}
\emline{11.00}{29.00}{175}{30.00}{48.00}{176}
\emline{11.00}{24.00}{177}{30.00}{43.00}{178}
\emline{11.00}{19.00}{179}{30.00}{38.00}{180}
\emline{11.00}{14.00}{181}{30.00}{33.00}{182}
\emline{11.00}{9.00}{183}{30.00}{28.00}{184}
\emline{14.00}{7.00}{185}{30.00}{23.00}{186}
\emline{14.00}{2.00}{187}{30.00}{18.00}{188}
\emline{18.00}{1.00}{189}{30.00}{13.00}{190}
\emline{23.00}{1.00}{191}{30.00}{8.00}{192}
\emline{28.00}{1.00}{193}{30.00}{3.00}{194}
\emline{33.00}{5.00}{195}{47.00}{19.00}{196}
\emline{50.00}{22.00}{197}{71.00}{43.00}{198}
\emline{33.00}{11.00}{199}{47.00}{25.00}{200}
\emline{47.00}{25.00}{201}{47.00}{25.00}{202}
\emline{50.00}{27.00}{203}{65.00}{42.00}{204}
\emline{33.00}{16.00}{205}{47.00}{30.00}{206}
\emline{50.00}{32.00}{207}{59.00}{41.00}{208}
\emline{34.00}{22.00}{209}{47.00}{35.00}{210}
\emline{50.00}{37.00}{211}{53.00}{40.00}{212}
\emline{34.00}{27.00}{213}{45.00}{38.00}{214}
\emline{34.00}{32.00}{215}{38.00}{36.00}{216}
\emline{34.00}{41.00}{217}{47.00}{54.00}{218}
\emline{43.00}{41.00}{219}{47.00}{45.00}{220}
\emline{39.00}{41.00}{221}{47.00}{49.00}{222}
\emline{34.00}{46.00}{223}{41.00}{53.00}{224}
\emline{33.00}{51.00}{225}{36.00}{54.00}{226}
\emline{50.00}{52.00}{227}{52.00}{54.00}{228}
\emline{51.00}{48.00}{229}{57.00}{54.00}{230}
\emline{57.00}{44.00}{231}{67.00}{54.00}{232}
\emline{51.00}{43.00}{233}{62.00}{54.00}{234}
\emline{63.00}{45.00}{235}{72.00}{54.00}{236}
\emline{69.00}{46.00}{237}{77.00}{54.00}{238}
\emline{74.00}{46.00}{239}{78.00}{50.00}{240}
\emline{33.00}{0.00}{241}{47.00}{14.00}{242}
\emline{50.00}{16.00}{243}{77.00}{43.00}{244}
\emline{39.00}{0.00}{245}{47.00}{8.00}{246}
\emline{50.00}{11.00}{247}{77.00}{38.00}{248}
\emline{45.00}{0.00}{249}{47.00}{2.00}{250}
\emline{50.00}{5.00}{251}{77.00}{32.00}{252}
\emline{50.00}{0.00}{253}{77.00}{27.00}{254}
\emline{56.00}{0.00}{255}{77.00}{21.00}{256}
\emline{63.00}{0.00}{257}{77.00}{14.00}{258}
\emline{69.00}{0.00}{259}{78.00}{9.00}{260}
\emline{75.00}{0.00}{261}{78.00}{3.00}{262}
\emline{34.17}{56.33}{263}{34.83}{57.17}{264}
\emline{37.00}{56.83}{265}{37.83}{58.00}{266}
\emline{40.50}{57.50}{267}{41.83}{58.83}{268}
\emline{44.33}{58.00}{269}{45.67}{59.33}{270}
\emline{52.50}{58.67}{271}{54.17}{60.33}{272}
\emline{56.67}{59.17}{273}{58.00}{60.50}{274}
\emline{60.83}{59.50}{275}{62.50}{61.17}{276}
\emline{65.17}{59.83}{277}{67.00}{61.67}{278}
\emline{70.00}{60.17}{279}{72.00}{62.17}{280}
\emline{74.17}{60.50}{281}{75.83}{62.17}{282}
\emline{31.00}{53.00}{283}{32.00}{53.00}{284}
\emline{31.00}{51.00}{285}{32.00}{51.00}{286}
\emline{31.00}{49.00}{287}{32.00}{49.00}{288}
\emline{31.00}{47.00}{289}{32.00}{47.00}{290}
\emline{31.00}{45.00}{291}{32.00}{45.00}{292}
\emline{31.00}{43.00}{293}{32.00}{43.00}{294}
\emline{31.00}{41.00}{295}{32.00}{41.00}{296}
\emline{31.00}{39.00}{297}{32.00}{39.00}{298}
\emline{32.00}{37.00}{299}{32.00}{37.00}{300}
\emline{31.00}{37.00}{301}{32.00}{37.00}{302}
\emline{31.00}{35.00}{303}{32.00}{35.00}{304}
\emline{31.00}{33.00}{305}{32.00}{33.00}{306}
\emline{31.00}{31.00}{307}{32.00}{31.00}{308}
\emline{31.00}{29.00}{309}{32.00}{29.00}{310}
\emline{31.00}{27.00}{311}{32.00}{27.00}{312}
\emline{31.00}{25.00}{313}{32.00}{25.00}{314}
\emline{31.00}{23.00}{315}{32.00}{23.00}{316}
\emline{31.00}{21.00}{317}{32.00}{21.00}{318}
\emline{31.00}{19.00}{319}{32.00}{19.00}{320}
\emline{31.00}{17.00}{321}{32.00}{17.00}{322}
\emline{31.00}{15.00}{323}{32.00}{15.00}{324}
\emline{31.00}{13.00}{325}{32.00}{13.00}{326}
\emline{31.00}{11.00}{327}{32.00}{11.00}{328}
\emline{31.00}{9.00}{329}{32.00}{9.00}{330}
\emline{31.00}{7.00}{331}{32.00}{7.00}{332}
\emline{31.00}{5.00}{333}{32.00}{5.00}{334}
\emline{31.00}{3.00}{335}{32.00}{3.00}{336}
\emline{31.00}{1.00}{337}{32.00}{1.00}{338}
\put(41.67,76.50){\makebox(0,0)[lc]{$s$}}
\put(39.33,70.67){\makebox(0,0)[rb]{$2$}}
\put(40.83,63.83){\makebox(0,0)[lb]{$3/2$}}
\put(41.83,55.00){\makebox(0,0)[lc]{$1$}}
\put(39.17,47.00){\makebox(0,0)[rc]{$1/2$}}
\emline{48.67}{56.33}{339}{48.67}{57.83}{340}
\emline{40.33}{59.33}{341}{42.00}{59.33}{342}
\emline{42.67}{59.33}{343}{45.50}{59.33}{344}
\emline{47.33}{59.33}{345}{48.50}{59.33}{346}
\put(39.17,59.33){\makebox(0,0)[rb]{$5/4$}}
\put(8.67,40.67){\makebox(0,0)[rb]{$-1$}}
\put(15.00,40.50){\makebox(0,0)[cb]{$\alpha^\ast$}}
\emline{32.50}{6.00}{347}{35.50}{6.00}{348}
\emline{36.33}{6.00}{349}{39.17}{6.00}{350}
\put(40.67,6.00){\makebox(0,0)[lt]{$s^\ast$}}
\put(41.83,10.00){\makebox(0,0)[lc]{$-2$}}
\put(41.83,25.00){\makebox(0,0)[lc]{$-1$}}
\put(41.33,35.00){\makebox(0,0)[lc]{$s^{\ast\ast}$}}
\put(29.83,41.00){\makebox(0,0)[rb]{$-1/3$}}
\put(38.83,41.00){\makebox(0,0)[rb]{$O$}}
\put(47.00,41.00){\makebox(0,0)[rb]{$1/3$}}
\put(78.00,41.67){\makebox(0,0)[cb]{$\alpha$}}
\put(54.00,43.50){\makebox(0,0)[lb]{$\bar k^2=0$}}
\bezier{108}(48.17,58.33)(63.83,60.17)(75.50,60.50)
\put(48.67,58.83){\circle{1.67}}
\end{picture}

\caption{\small Diagram of $A$-stability of FRW-models. Inclined dash
points  region of  $A$-instability of FRW-models. The boundary of this region
is double line.  At this boundary
FRW-models can be  $A$-stable (vertical line  $\alpha=1/3,$ horizontal line $s=1$
and parts of rational dependency
(\ref{rell}) with end at the point $\{\alpha^\ast;s^\ast\}$
and the beginnings at the  points $\{-1/3;s^{\ast\ast}\}$
and $\{-1;2\}$ (formula (\ref{bounce})),
conditionally  $A$-stable (vertical line (with horizontal dash) $\alpha=-1/3$
under (\ref{part4}); whole  performed part of this line corresponds to open models) or
$A$-unstable (part (with inclined dash) of  the dependency (\ref{part2}), beginning at the point $\{-1/3;1\}$;
empty point on this branch has coordinates $\{1/3;5/4\}$ and corresponds to intersection
of (\ref{part2}) with vertical line  $\alpha=1/3,$ where  all models
are stable).
$A$-stable branch, which asymptotically tends to line $s=1/2,$
corresponds to $\bar k=0.$
}\label{d2}
\end{figure}
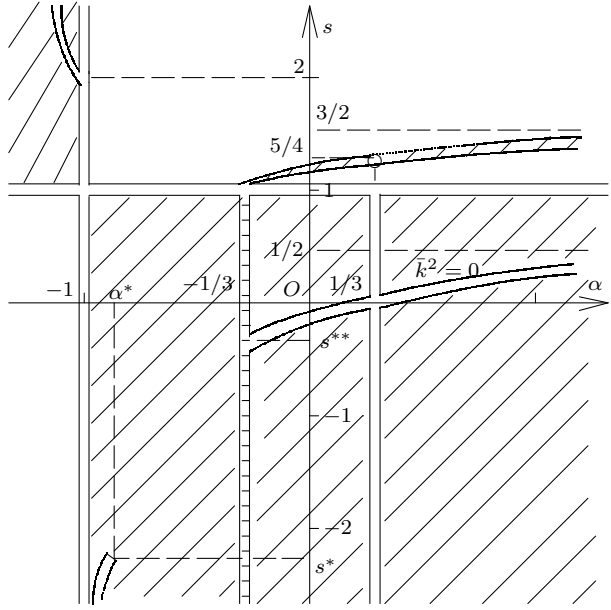

All results are summarized in diagram of   $A$-stability (fig. \ref{d2}).

\subsection{Structural stability of FRW-models with nonlinear matter}

Let consider now the equation  (\ref{ggen1}) with right-hand side of kind
$x^{-3(\alpha+1)(b+1)+3},$ responsible for perturbation by nonlinear
properties of matter.
In the limits $x\to0$ under $1+3\alpha>0$
this equation can be reduced to the form:
\begin{equation}\label{en1}
2x^2(x\xi(x))''+(1-3\alpha)x(\xi(x)-(x\xi(x))')=
x^{-3(1+\alpha)(b+1)+4+3\alpha},
\end{equation}
where $\xi(x)=-\delta(x)/\alpha\sigma c^{b+1}x.$
Its particular solution:
\begin{equation}\label{n1}
\xi(x)|_{x\to0}=\frac{x^{-3b(1+\alpha)}}{3b(-3\alpha+6\alpha b-1+6b)(1+\alpha)}.
\end{equation}
Simple analysis of index shows, that {\it almost all
FRW-models with nonlinear matter
are $S$-stable under $b<0,$ and are $S$-unstable under $b>0$}
(see also interesting analysis in \cite{quint,odin3})).

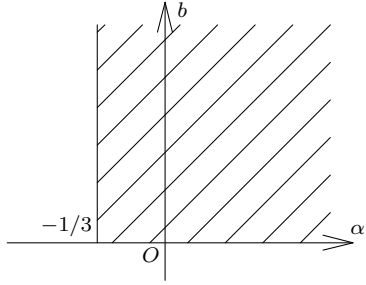
\begin{figure}[htb]
\footnotesize
\centering \unitlength=0.50mm \special{em:linewidth 0.4pt}
\linethickness{0.4pt}
\unitlength=1mm
\special{em:linewidth 0.4pt}
\linethickness{0.4pt}
\begin{picture}(48.00,39.00)
\emline{2.00}{7.00}{1}{48.00}{7.00}{2}
\emline{48.00}{7.00}{3}{44.00}{8.00}{4}
\emline{44.00}{6.00}{5}{48.00}{7.00}{6}
\emline{23.00}{2.00}{7}{23.00}{39.00}{8}
\emline{23.00}{39.00}{9}{22.00}{35.00}{10}
\emline{24.00}{35.00}{11}{23.00}{39.00}{12}
\emline{14.00}{36.00}{13}{14.00}{7.00}{14}
\emline{14.00}{30.00}{15}{20.00}{36.00}{16}
\emline{14.00}{25.00}{17}{25.00}{36.00}{18}
\emline{14.00}{20.00}{19}{30.00}{36.00}{20}
\emline{14.00}{15.00}{21}{35.00}{36.00}{22}
\emline{14.00}{10.00}{23}{40.00}{36.00}{24}
\emline{16.00}{7.00}{25}{45.00}{36.00}{26}
\emline{21.00}{7.00}{27}{45.00}{31.00}{28}
\emline{26.00}{7.00}{29}{45.00}{26.00}{30}
\emline{31.00}{7.00}{31}{45.00}{21.00}{32}
\emline{36.00}{7.00}{33}{45.00}{16.00}{34}
\emline{41.00}{7.00}{35}{45.00}{11.00}{36}
\emline{14.00}{35.00}{37}{15.00}{36.00}{38}
\put(47.67,8.17){\makebox(0,0)[lb]{$\alpha$}}
\put(24.67,38.00){\makebox(0,0)[lc]{$b$}}
\put(13.33,7.83){\makebox(0,0)[rb]{$-1/3$}}
\put(22.33,6.33){\makebox(0,0)[rt]{$O$}}
\end{picture}

\caption{\small Diagram of  $S$-stability of  FRW-models with generalized nonlinear equation of state.
Inclined dash shows region of $S$-instability of FRW-models.
Boundary belongs to the region.
}\label{d3}
\end{figure}

Particular solution to the asymptotic equation (\ref{en1})
under $b=0$ has the form:
\begin{equation}\label{n2}
\xi(x)=\frac{(3\alpha+1)\ln x-2}{(1+3\alpha)^2}.
\end{equation}
We see, that such models are
$S$-unstable.
Particular solution to the equation (\ref{en1}) under $b=(1+3\alpha)/6(1+\alpha),$
when denominator of (\ref{n1}) vanishes, takes the form:
\begin{equation}\label{n3}
\xi(x)|_{x\to0}=-\frac{(3\alpha+1)\ln x+2}{(1+3\alpha)^2}x^{-(1+3\alpha)/2},
\end{equation}
and we see, that such models are $S$-instable too.
Results of our analysis are shown in diagram (fig. \ref{d3}).

In case  $(1+3\alpha)<0$ analysis of $A$-stability
is reduced to previous formulas of this section, if one
will consider their
behaviour under $x\to\infty.$ So, from the expression (\ref{n1})
it follows, that {\it almost all FRW-models with $-1<\alpha<-1/3$ are $A$-stable under  $b>0$
and are $A$-unstable under  $b<0$; almost all FRW-models with $\alpha<-1$ are
$A$-stable under  $b<0$
and are $A$-unstable under $b>0.$} Formulas (\ref{n2})-(\ref{n3}) allow
us to conclude, that in all particular cases, which have been considered above,
the models are
$A$-unstable.

For the case $1+3\alpha>0$ we obtain the following asymptotic form of equation  (\ref{ggen1}):
\begin{equation}\label{en2}
-2\bar k^2 x^2(x\xi(x))''+6\alpha\bar k^2x(\xi(x)-(x\xi(x))')=
x^{-3(1+\alpha)(b+1)+3}.
\end{equation}
Its particular solution has the form:
\begin{equation}\label{n4}
\xi(x)|_{x\to\infty}=-\frac{x^{-1-3\alpha b-3\alpha-3b}}{6b(1+\alpha)\bar k^2(3\alpha b+3\alpha+3b+1)}.
\end{equation}
Analysis of index shows, that
{\it almost all FRW-models with strong energodominancy condition
are $A$-stable under
\begin{equation}\label{boundd}
b<-(1+3\alpha)/3(1+\alpha)
\end{equation}
and are  $A$-unstable in case of opposite inequality.}

Further  analysis of particular cases of the models with strong energodominancy condition reveals
that:
\begin{itemize}
\item
under  $b=0$ models are $A$-stable;
\item
under $b=-(1+3\alpha)/3(1+\alpha)$  models are $A$-unstable;
\item
under $\bar k=0$  models are $A$-stable under  $b>0,$  while under $b\le0$  models are
 $A$-unstable;
\item
under $\bar k=0$ and $\alpha=-(6b-1)/3(2b-1)$  models are $A$-stable;
\item
under $\alpha=-1/3$  models are $A$-stable and $S$-unstable when  $b>0,$  $A$-unstable  and
$S$-stable when $b<0.$ Under $b=0$  models are unstable in all senses.
\item
under $\alpha=-1$ models are  $A$-unstable under any value of $b.$
\end{itemize}

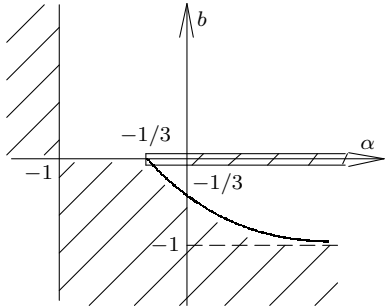
\begin{figure}[htb]
\footnotesize
\centering \unitlength=0.50mm \special{em:linewidth 0.4pt}
\linethickness{0.4pt}
\unitlength=1mm
\special{em:linewidth 0.4pt}
\linethickness{0.4pt}
\begin{picture}(51.17,40.17)
\emline{25.00}{40.00}{1}{25.00}{0.00}{2}
\emline{1.67}{19.50}{3}{51.17}{19.50}{4}
\emline{51.17}{19.50}{5}{46.50}{20.33}{6}
\emline{46.50}{18.50}{7}{51.17}{19.50}{8}
\emline{24.00}{35.67}{9}{25.00}{40.17}{10}
\emline{25.00}{40.17}{11}{25.83}{35.67}{12}
\emline{8.00}{40.00}{13}{8.00}{0.67}{14}
\bezier{116}(19.67,19.50)(29.33,8.67)(43.67,8.50)
\emline{25.00}{20.17}{15}{46.00}{20.17}{16}
\emline{25.00}{18.67}{17}{46.00}{18.67}{18}
\emline{1.00}{20.00}{19}{8.00}{27.00}{20}
\emline{1.00}{25.00}{21}{8.00}{32.00}{22}
\emline{1.00}{30.00}{23}{8.00}{37.00}{24}
\emline{1.00}{35.00}{25}{6.00}{40.00}{26}
\emline{6.00}{20.00}{27}{8.00}{22.00}{28}
\emline{25.67}{18.67}{29}{27.17}{20.17}{30}
\emline{30.50}{18.67}{31}{32.17}{20.17}{32}
\emline{35.67}{18.67}{33}{37.17}{20.17}{34}
\emline{41.17}{18.67}{35}{42.33}{20.17}{36}
\emline{45.67}{18.83}{37}{46.33}{20.17}{38}
\put(26.33,38.17){\makebox(0,0)[lc]{$b$}}
\put(49.00,20.83){\makebox(0,0)[cb]{$\alpha$}}
\emline{25.00}{20.17}{39}{19.50}{20.17}{40}
\emline{19.50}{20.17}{41}{19.50}{18.67}{42}
\emline{19.50}{18.67}{43}{25.00}{18.67}{44}
\emline{8.00}{17.00}{45}{10.00}{19.00}{46}
\emline{8.00}{12.00}{47}{15.00}{19.00}{48}
\emline{8.00}{7.00}{49}{19.00}{18.00}{50}
\emline{8.00}{2.00}{51}{22.00}{16.00}{52}
\emline{12.00}{0.00}{53}{25.00}{13.00}{54}
\emline{18.00}{0.00}{55}{29.00}{11.00}{56}
\emline{24.00}{0.00}{57}{33.00}{9.00}{58}
\emline{30.00}{0.00}{59}{38.00}{8.00}{60}
\emline{36.00}{0.00}{61}{44.00}{8.00}{62}
\emline{42.00}{0.00}{63}{45.00}{3.00}{64}
\put(19.50,21.17){\makebox(0,0)[cb]{$-1/3$}}
\put(7.17,18.67){\makebox(0,0)[rt]{$-1$}}
\put(25.67,14.83){\makebox(0,0)[lb]{$-1/3$}}
\emline{25.00}{8.00}{65}{27.00}{8.00}{66}
\emline{28.00}{8.00}{67}{30.00}{8.00}{68}
\emline{31.00}{8.00}{69}{33.00}{8.00}{70}
\emline{34.00}{8.00}{71}{36.00}{8.00}{72}
\emline{37.00}{8.00}{73}{39.00}{8.00}{74}
\emline{40.00}{8.00}{75}{42.00}{8.00}{76}
\emline{43.00}{8.00}{77}{45.00}{8.00}{78}
\put(24.00,8.00){\makebox(0,0)[rc]{$-1$}}
\end{picture}
\caption{\small Diagram of $A$-stability of  FRW-models with generalized nonlinear
equation of state. Inclined dash shows region, where the models are  $A$-unstable.
Boundary of this region belongs to the region.
Main part of this region on lower half-plane is bounded by the curve
(\ref{boundd}). Half-line $b=0,$ $\alpha>-1/3$
corresponds to flat $A$-unstable FRW-models.
}\label{d4}
\end{figure}

All results are summarized in diagram of  $A$-stability (fig. \ref{d4}).

\section{Stability of standard models with $\Lambda$-term}\label{lambda}

So far we have considered non-perturbed FRW-models
with $\Lambda=0$.
One may hope, that extension of class of non-perturbed models
will lead to the extension of their stability region.
Let us illustrate this idea by example of non-perturbed FRW-models with non-zero
$\Lambda$-term. For our purposes it will be sufficient to restrict
ourselves
by FRW-models with dust-like matter ($\alpha=0$) for generalized nonlinear theories
of gravity and by the FRW-models with isotropic  radiation
($\alpha=1/3$) for generalized models with non-linear matter.
By note, made after formula
(\ref{solg}), inclusion of  $\Lambda$-term does not change
behaviour of FRW-models near singularity, that may be checked directly
by our method.
The diagram  \ref{d2} implies, that generalized FRW-models with high orders of scalar
curvature кривизне членами under $\alpha=0$
(and under $\Lambda=0$) are $A$-stable under $s\in[1;\infty)\setminus\{7/6\}$ for all
values of
$\bar k^2\le0$ and under $s=(7-\sqrt{73})/12\approx-0.13$ for $\bar k=0.$
The diagram  \ref{d4} implies, that generalized FRW-models with isotropic radiation,
perturbed by non-linearity of state equation
(under $\Lambda=0$), are $A$-stable under $b\in[-1/2;\infty)$ for all values of  $\bar k^2\le0,$
excepting the value $b=0$ under $\bar k=0.$

Standard calculations with using formulas (\ref{eq20})-(\ref{ff3}), taken under
$\Lambda\neq0,$
lead to the following kind of equation on structural perturbations for models
with non-linear gravity under $\alpha=0$ in the limit
$x\to\infty:$
\[
x^5\delta''+x^4\delta'-4x^3\delta\sim x^4,
\]
where we have omitted  constant factor in the right-hand side.
Its particular solution: $\xi=\delta/x\sim\text{const},$
that means  {\it unconditional $A$-stability of all FRW-models with  $\Lambda$-term
and dust-like matter.}

For FRW-models with $\Lambda$-term and nonlinear equation of state
we obtain equation for structural perturbation of the following kind:
\[
x^4\delta''+2x^3\delta'-6x^2\delta\sim x^{-1-4b}.
\]
Its particular solution:
$
\xi(x)\sim x^{-4(1+b)}.
$
So, we conclude, that  {\it due to inclusion of  $\Lambda$-term, region of  $A$-stability of
FRW-models with isotropic radiation become larger:
it includes now additional segment of  $b\in[-1;-1/2]$
and the value $b=0.$ }
This conclusion is closely related to results of papers
\cite{sh1,sh2},
where generic structural stability of $\Lambda$CDM cosmological
models has been established by analysis of topological properties of
phase portraits of cosmological models.

Complete analysis of structural stability of FRW-models with
$\Lambda$-term we defer for future publications.

\section{Conclusion}

We have shown, that the question
of structural stability of standard cosmological FRW-models
is of  non-trivial nature and can be investigated in general manner
in frame of outlined classes of generalized cosmological models,
including non-linear gravity and non-linear properties of matter.
We have restricted ourselves to investigations of structural stability
near the cosmological singularity
under  $a(t)\to0$ ($S$-stability)
and under $a(t)\to\infty$ ($A$-stability).
Note, that  in terms of relative
structural perturbation $S$-stability condition  takes the form:
\begin{equation}\label{steq}
\lim\limits_{a\to0}\frac{\delta}{a}<\infty.
\end{equation}
So, $S$-instability
condition can be considered in the following two qualitatively different
situations.
As it can be seen from the content of sections \ref{ge} and \ref{lambda},
in almost  all cases asymptotic behaviour  of structural perturbation
is described by power function of
$a$: $\delta\sim a^p.$ The condition (\ref{steq})
leads then to the inequality  $p\ge1,$ providing $S$-stability.
It is obvious, that under $0\le p<1$ {\it singularity is conserved,
but behaviour of scale factor near singularity  changes.} In case  $p<0,$
{\it singularity is destroyed.}
In our diagrams of  $S$-stability  these situations are indistinguishable.
In case of  $A$-instability we, obviously, always deal with change of
asymptotic behaviour of scale factor.

Union of all stability diagrams, which corresponds to simultaneous action
of structural perturbations of all considered types, leads to the following
conclusion:
{\it only the models with $-1<\alpha<-1/3,$ $b>0,$ $s\ge1$ and the
 models with $\alpha<-1,$ $b<0,$ $s\le1$ together with  some particular families of cosmological models,
  having zero measures on diagrams
\ref{d1}-\ref{d2}, possess complete (in frame of chosen
class of perturbations) structural stability.}
It, in particular, means that {\it almost all  FRW-models
with strong energy condition are structurally unstable.}

In our opinion the question of structural stability in  cosmology
is of principal significance. Consequent account of new observational data
and evolution of our theoretical tools of descriptions of nature
lead to unavoidable theoretical modifications of basic cosmological equations.
If we don't speak about scientific revolution and about change of paradigm,
such modifications will be looked at as, in some sense, small corrections
in lagrangians or in equations.
Our analysis shows that in spite of "smallness"\,,
this corrections may lead to the  total deleting of one of the properties of original theory,
to significant modifications of others properties  and to insignificant quantitative
variations of the third ones.
In the first two cases the properties of original models are
"thin"\, with respect to the modifications of the theory and so, in the context of generalized
models, they are "accidental".
Our investigation reveals, that cosmological singularity is
such "accidental"\, property of Einstein cosmology.
Any non-linear corrections
of kind $\lambda R^s$ under $s\ge2$ in gravitational lagrangian
will destroy cosmological singularity, whatever  small
constant $\lambda$ is chosen. In fact, cosmological singularity is not experimentally observed.
It is desirable {\it to provide stability of  those properties of cosmological models
which have robust experimental (observational) support}.
In spite of infinity (and even uncountability) of a set of possible generalizations
of standard cosmology, the set of its {\it viable and reasonable} (from the viewpoint of present time) generalizations
is quite foreseeable and can be analyzed from general positions
 \cite{peldan}.

Our example with $\Lambda$-term
shows that among the set of allowed models one can try to find
relatively simple one, which possesses sufficient "structural security"\,
and conserve some important properties under its possible further modifications.
Note, that considerations of structural stability can be viewed as purely theoretical argument for
inclusion of $\Lambda$-term into Einstein theory of gravity.
Lets note also that diagrams \ref{d2} and \ref{d4} throw light on the
following question: what class of generalized cosmological models can be
relevant to accelerated in latter times universe? The answer is
the following: {\it for FRW models with strong energodominancy condition $(1+3\alpha)>0$ such
models must be $A$-unstable!} In other words relevant generalized
models must have in their lagrangians terms  of the kind $R^s$ with $s<1$ and
(or) equation of state with nonlinearity index $b<-(1+3\alpha)/3(1+\alpha).$

Using technic of conform transformations
\cite{others}
our considerations can be directly applied to the cosmological models within
Einstein gravity with self-interacting
scalar fields and within  non-riemannian geometrical theories.
This topic is far from the scope of present paper.

In fact, the question of structural stability is not limited only to cosmology.
It is valid in all situations when physics is local and can be
described by sets of differential equations. By general considerations
any local physical model can be "slightly"\, modified, such that any forgiven property
of the model will be destroyed.
Probably, in this situation some fundamental principles will be  helpful to restrict
the set of possible modifications of the model and to "stabilize"\,  some of its important
physical properties.

\end{document}